\documentclass[eqsecnum,twocolumn,aps]{revtex4}

\usepackage{graphicx}%
\usepackage{dcolumn}
\usepackage{amsmath}

\begin{document}

\preprint{HEP/123-qed}

\title[Short Title]{Surface effects in nucleation and growth of 
smectic B crystals in thin samples}

\author{T. B\"orzs\"onyi $^{1,2}$ and S. Akamatsu$^1$}
  \email{akamatsu@gps.jussieu.fr}
\affiliation{$^1$Groupe de Physique des
Solides, CNRS UMR 7588, Universit\'es Denis-Diderot et
Pierre-et-Marie-Curie, Tour 23, 2 place Jussieu, 75251 Paris Cedex 05,
France\\
$^2$Research Institute for Solid State Physics and Optics,
  Hungarian Academy of Sciences,  H-1525 Budapest, P.O.B.49, Hungary
}

\date{\today}

\begin{abstract}
We present an experimental study of the surface effects (interactions
with the container walls) during the nucleation and growth of smectic
B crystals from the nematic in free growth and directional
solidification of a mesogenic molecule ($\rm{C_4H_9-(C_6H_{10})_2CN}$)
called CCH4 in thin (of thickness in the 10~$\mu$m range) samples.  We
follow the dynamics of the system in real time with a polarizing
microscope.  The inner surfaces of the glass-plate samples are coated
with polymeric films, either rubbed polyimid (PI) films or
monooriented poly(tetrafluoroethylene) (PTFE) films deposited by
friction at high temperature.  The orientation of the nematic and the
smectic B is planar.  In PI-coated samples, the orientation effect of
SmB crystals is mediated by the nematic, whereas, in PTFE-coated
samples, it results from a homoepitaxy phenomenon occurring for two
degenerate orientations.  A recrystallization phenomenon partly
destroys the initial distribution of crystal orientations.  In
directional solidification of polycrystals in PTFE-coated samples, a
particular dynamics of faceted grain boundary grooves is at the origin
of a dynamical mechanism of grain selection.  Surface effects also are
responsible for the nucleation of misoriented terraces on facets and
the generation of lattice defects in the solid.
\end{abstract}

\pacs{PACS numbers: 64.70.Md, 81.10.Aj, 64.70.Dv, 68.70.1w}


\maketitle


\section{Introduction}

\begin{figure}[htbp]
\includegraphics[width=7cm]{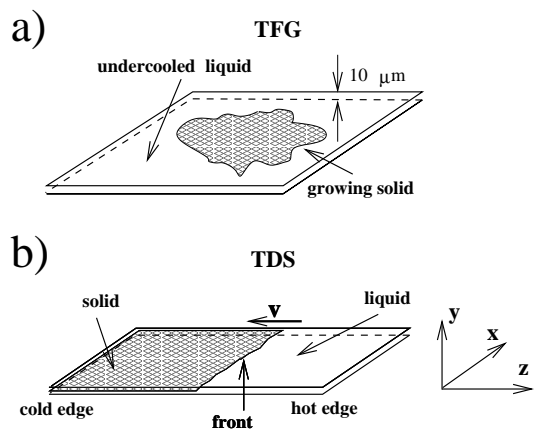}
\caption{Principle of thin-sample solidification experiments: a) TFG: 
thin-sample free growth; b) TDS: thin-sample 
directional-solidification.  $z$: axis of the thermal gradient $G$.  
$x$: axis parallel to the isotherms.  $y$ : transverse direction.  
$V$: pulling velocity.}
\label{sketches}
\end{figure}

The appearance of molecular crystals in a supercooled liquid occurs
generally by a heterogeneous-nucleation process.  In many cases, the
walls of the container play the role of preferential nucleation
substrate.  Depending on the nature of that substrate, crystals may
grow in epitaxy with it, and their orientation be well controlled.  It
has been known for a long time that one can also orient mesophases in
thin (of thickness in the $10~\mu \rm{m}$ range) samples of a
mesogenic molecule by coating the inner surfaces of the container,
generally made of glass, with a molecularly thin film of suitable
nature and topography \cite{cognard}.  For nematic and smectic A
phases, the microscopic mechanism of phase orientation at play is a
combination of surface energy minimization and elastic effects
specific of the short-range orientational order proper to those phases
\cite{degennes}.  Recently, surface coatings promoting both the
alignment of a mesophase and the selection of the orientation of a
smectic B (SmB) phase in coexistence with it have been used in free
growth (solidification in a uniformly undercooled sample)
\cite{tobo96,gora96,toeb99,boto00,review} and directional
solidification (solidification at a constant speed $V$ along a fixed
temperature gradient $G$) \cite{osme89,meos91,boak01} of different
mesogenic molecules in thin samples (Figure \ref{sketches}).  A SmB
phase is a true crystal with long-range positional order in the three
directions of space --it is not a mesophase, but a lamellar plastic
crystal.  Thanks to the possibility of controlling the orientation of
the two coexisting phases, a variety of new stationary, faceted growth
patterns, resulting from a complex combination of a diffusion
controlled dynamics and of a non-linear growth kinetics proper to facet
orientations, has been discovered \cite{boak01}.  However, some
effects of the interactions between the molecules and the walls of the
container (surface effects) on the growth dynamics of the crystal
remain to be studied.

In this paper, we present an experimental study of surface effects
during the nucleation and growth of SmB crystals from the nematic
phase of the mesogenic molecule $\rm{C_4H_9-(C_6H_{10})_2CN}$ (in
short, CCH4).  The inner surfaces of our glass-wall samples are coated
with molecularly thin polymer films, either rubbed polyimid (PI)
films \cite{cognard,feller} or monooriented poly(tetrafluoroethylene)
(PTFE) films deposited by friction transfer at high temperature
\cite{tabor,wism91,schott94,ueda,dado97}.  The CCH4 substance is a
member of a series of mesogenic molecules, noted CCHm, where $m$~=~3
to 5 is the length of the aliphatic chain of the amphiphilic liquid
crystal, which undergo a first-order transition from the nematic to
the SmB phase at a temperature $T_{NS}$ which depends slightly on $m$
($T_{NS} \approx 53^oC$ for CCH4) \cite{lead}.  We used both the
thin-sample free growth (TFG) and thin-sample directional
solidification (TDS) methods to observe the time evolution of the
shape of the solid-liquid interface with an optical microscope.  The
practically two-dimensional (2D) character of the samples implies that
the solid-liquid interface remains essentially perpendicular to the
sample plane.  In the situations considered in the present study,
there is no convection in the liquid, and matter exchanges occur only
by diffusion.

\begin{figure}[htbp]
\includegraphics[width=8cm]{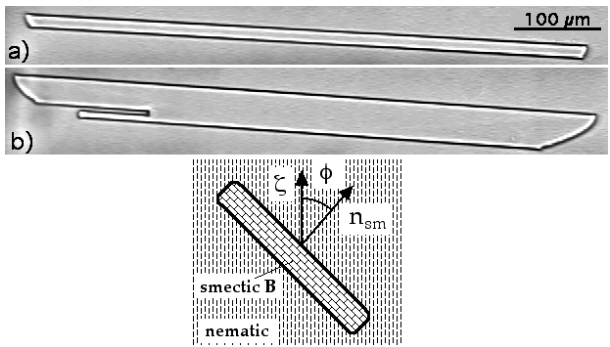}
\caption{Thin-sample free growth of a smectic B crystal of CCH4 in a
10-$\mu \rm{m}$ thick PI-coated sample (the rubbing axis $\zeta$ is
vertical).  No polars.  a) $\Delta T = 0.07$ K. The facets do not
grow.  b) $\Delta T = 0.11$ K. The facets grow, and the tips undergo a
morphological instability.  Bottom: definition of the disorientation
angle $\phi$ and the vector normal to the smectic layers
${\bf{n}}_{sm}$.}
\label{sticks}
\end{figure}

Near equilibrium, SmB crystals in coexistence with the nematic exhibit
a single facet plane, namely, the smectic-layer plane (Figure
\ref{sticks}) \cite {tobo96}.  The nematic-SmB interface is otherwise
rough on a molecular scale.  In PI- and PTFE-coated samples, a planar
orientation (see below) is imposed not only to the nematic, but also
to the SmB phase.  In freshly filled samples, i.e., CCH4 nematic
samples in which no crystallization has yet occurred, the nematic is
aligned along the direction $\zeta$ of rubbing (for PI) or friction
(for PTFE) of the polymer film.  For a planar orientation of the SmB
phase, the smectic layers, thus the facets of the nematic-SmB
interface, are perpendicular to the sample plane, so that the 2D
character of the system is guaranteed.  The partly faceted crystals
then grow in a well oriented nematic which presents large regions free
of defects, which, if present, would perturb the interface.

Most previous experimental studies of the growth dynamics of CCHm SmB
crystals were performed in TFG in PI-coated samples
\cite{tobo96,gora96,toeb99,boto00,review}.  In a more recent TDS
study, thin CCH4 PTFE-coated samples were used for the first time
\cite{boak01}, in order to increase the strength of the selection of
the in-plane orientation of SmB crystals.  The purpose of the present
paper is to cast light to the mechanisms at play in that selection.
Therefore, we will focus our attention on the formation by
heterogeneous nucleation and growth, and the coarsening of SmB
polycrystals of CCH4 in TFG, and on the growth dynamics of such
polycrystals in TDS. We reproduced some TFG experiments in PI-coated
samples.  We thus observed some unexpected phenomena, such as the
rotation of highly misoriented crystals (see below) during the first
stages of their growth.  However, the most interesting results have
been obtained in PTFE-coated samples, for reasons which will become
clearer later on.

Our main TFG results can be summed up as follows.  In both PI- and 
PTFE-coated samples, many CCH4 SmB crystals of a planar orientation 
nucleate for undercoolings of a few 0.1 K. Let us define the 
(in-plane) disorientation angle $\phi$ as the angle between $\zeta$ 
and the unit vector ${\bf{n}}_{sm}$ normal to the smectic layers (Fig.  
\ref{sticks}).  The average value of $\phi$ is equal to 0 in a 
PI-coated sample, but the width of its distribution about 0 is large 
($\approx 60\rm{^o}$).  This weak orientation selection effect and the 
above-mentioned process of rotation of highly misoriented crystals 
(i.e., of large $\phi$ values) suggest that the final in-plane 
orientation of SmB crystals in PI-coated samples is determined by the 
interactions of the crystals with the nematic.  In contrast, in 
PTFE-coated samples, SmB crystals directly nucleate with either of two 
symmetrical disorientations $\pm \phi_{ptfe}$ (about $\pm 13 
\rm{^o}$).  This apparent epitaxial growth suggests the existence of a 
specific, strongly anisotropic interaction between the CCH4 molecules 
and the PTFE film on a molecular scale.  This is supported by the 
existence of a strong "memory effect" \cite{clark, ouchi} in re-melted 
PTFE-coated samples.  Such an effect is almost absent in PI-coated 
ones.

\begin{figure}[htbp]
\includegraphics[width=8cm]{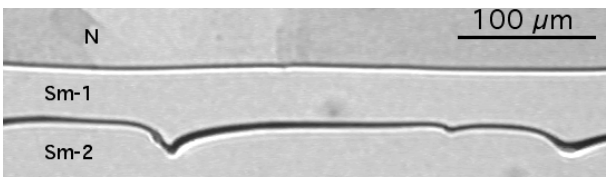}
\caption{Planar-front regime in thin-sample directional solidification of
CCH4 in a PTFE-coated sample ($G = 54 ~{ \rm{Kcm^{-1}}}$; $V=0.9~\mu{
\rm{ms^{-1}}}$).  The friction axis is vertical.  In this, and all the
following micrographs, growth is upwards.  N: nematic.  Sm-1: smectic B.
Sm-2: smectic B oriented differently from Sm-1.  Note the domains in the
nematic.  Sm-1 is a single crystal, but Sm-2 a polycrystal, as shown by the
presence of cusps caused by grain boundaries on the interface between 
Sm-1 and Sm-2 (recrystallization front).}
\label{planarfront}
\end{figure}

In TDS of single crystals of CCH4 \cite{boak01}, the front is 
generally (i.e., except for very special orientations) non-faceted, 
and exhibits a dynamics similar to that of any nonfaceted crystal at 
small solidification rates.  The front remains planar below a 
threshold velocity $V_{c}$ (Figure \ref{planarfront}) and becomes 
cellular above $V_{c}$.  Facets appear only above $V_{c}$, which 
causes the formation of localized objects, comparable to solitary 
waves, called "facetons".  In the case of polycrystal samples, that we 
consider in the present study, facets appear even for $V < V_{c}$ in 
the vicinity of grain boundaries (GB's).  This has many consequences 
in PTFE-coated samples, the most remarkable of which is the existence 
of a mechanism of grain selection, the main ingredient of which are 
the particular dynamics of faceted grooves attached to GB's and the 
nucleation of SmB crystals ahead of the front.  We study that 
mechanism in detail.  We also consider the existence of a 
recrystallization front visible in the rear of the solidification 
front.  Finally, a careful analysis of the stepwise growth dynamics of 
the facets allows us to identify a mechanism of generation of planar 
lattice defects in the solid.

\section{Experimental Section}
\subsection{Preparation of the samples}
The basic thermodynamical parameters of CCH4 (Merck) and the other 
CCHm compounds can be found in refs.  \cite{tobo96,boak01}.  The 
crystal parameters of the SmB phase of CCH4 were measured by 
small-angle X-ray scattering in a previous study \cite{boak01} (see 
ref.  \cite{lead} for those of CCH3 and CCH5).  The lamellar stacking 
of the molecules in the SmB is of the AB type for all CCHm compounds.  
The packing of the molecules within the layers is hexagonal.  The 
phenomena of selection of crystal orientation in PI- and PTFE -coated 
samples do not depend qualitatively upon whether CCH3, CCH4 or CCH5 
are considered.  Some quantitative differences will be mentioned later 
on.  As received, CCH4 contains a small amount of unknown impurities.  
A rough characterization of the residual impurities can be found in 
ref.  \cite {boak01} --the thermal gap $\Delta T_{0}$ is about 0.2 K, 
the partition coefficient 0.12, and the diffusion coefficient in the 
nematic $8 \times 10^{-7}~{\rm{cm^2 s^{-1}}}$.  Because of a chemical 
decomposition taking place in the nematic phase, the impurity 
concentration in a given sample increases slowly in time, which shows 
up by a progressive decrease of the nematic-SmB equilibrium 
temperature $T_{NS}$ ($T_{NS}$ actually is the temperature of the 
liquidus of the alloy CCH4 + residual impurities).  This is, however, 
of secondary importance for the present purpose.

The PI-coated cells (thickness $d = 10~\mu \rm{m}$; lateral dimensions of
$12 \times 20~\rm{mm^2}$) were purchased from E.H.C Co., Japan.  The
PTFE-coated cells ($d = 12~\mu\rm{m}$; lateral dimensions of $9 \times
60~\rm{mm^2}$) were made in our laboratory.  Monooriented PTFE films are
deposited by slowly sliding a PTFE block pressed against the surface of a
clean glass microslide maintained at a temperature slightly higher than
250~{$\rm{^o}$}C, along a direction $\zeta$.  No further treatment is
applied.  Two PTFE-coated plates, separated from each other by two parallel
12-$\mu \rm{m}$ thick plastic spacers, are assembled and glued so as to
make a thin cell.

We filled our samples by capillarity, according to either of two different
procedures.  A first method (method 1), used in TFG only, consists of
filling the sample {\it{in~situ}} at a temperature higher than the
isotropic-nematic equilibrium temperature $T_{IN}$ (about 80 {$\rm{^o}$}C).
The sample is then used without being sealed or outgased.  In the second
method (method 2), a sample is filled under an Ar atmosphere at a
temperature higher than $T_{NS}$, then rapidly cooled down to room
temperature, and sealed.  It is then a SmB polycrystal, as a result of the
heterogeneous nucleation of many crystals in the nematic.

\subsection{Nematic alignment}
When a nematic phase is in contact with a flat homogeneous wall the 
orientation of the director {$\bf{d}$} is an intermediate between two 
particular configurations, or anchorings, called "homeotropic" and 
"planar", corresponding to {$\bf{d}$} being perpendicular and parallel 
to the wall, respectively \cite{degennes}.  The surface induced order 
propagates over a macroscopic distance (persistence length) of order 1 
$\mu \rm{m}$ into the bulk because of the particular elastic 
properties of the nematic.  This allows one to obtain a uniformly 
aligned nematic when the sample is sufficiently thin.  That a uniform 
planar (or almost planar \cite{pretilt}) alignment along a pre-defined 
direction $\zeta$ is obtained thanks to a gentle rubbing of a PI film 
with a soft textile brush is a well-known empirical fact.  The 
physical origin (influence of a one-dimensional microscopic roughness, 
anisotropic and/or specific molecular interactions) of that effect 
still remains controversial \cite{degennes,feller,vegt}.

Monooriented PTFE films have been known for a long time, at least 
empirically, to promote epitaxial growth of molecular crystals of 
various organic compounds \cite{tabor,wism91, schott94,ueda,dado97}.  
Recently, their structure has been revealed by electron diffraction 
\cite{ueda}; they are almost fully crystalline (as bulk PTFE is), 
whereas PI films are generally partly crystalline.  A lattice matching 
between PTFE films and molecular crystals has been evidenced 
experimentally in a few cases \cite{wism91}.  On the other hand, 
atomic-force microscopy studies revealed that a monooriented PTFE film 
deposited at a temperature above 250 {$\rm{^o}$}C (as in the present 
study) onto a flat silicon wafer under well controlled conditions of 
temperature, pressure and sliding speed, is not of uniform thickness, 
but always contains thin stripes of a typical thickness of 10 nm and a 
width of several 100 nm, lying parallel to the friction axis $\zeta$ 
\cite{schott94,dado97}.  These stripes, which are probably made of 
well-crystallized bunches of PTFE chains, remain perfectly rectilinear 
and uninterrupted along millimetric distances.  This uniaxial 
roughness probably plays a major role in the alignment of the nematic 
along $\zeta$ \cite {gold}.

In PI-coated samples, the alignment of the nematic along the rubbing axis
is generally uniform, independently of the filling method (see below).  In
PTFE-coated samples filled according to method 1, and cooled down to a
temperature slightly above $T_{NS}$, nematic regions with the expected
average alignment along $\zeta$ delimited by more or less extended defect
zones are observed.  In the well-aligned regions, a faintly contrasted
striation parallel to $\zeta$ is visible between crossed polars (see, for
instance, Figure \ref{nuclPTFE} below).  The defects between the aligned
regions are clearly associated with imperfections of the PTFE film (small
aggregates of the PTFE chains).  The striation within the relatively
uniform regions are probably due to exceptionally thick, but well
crystallized, bunches of polymer chains.  When prepared by method 2,
PTFE-coated samples are always structured, in the nematic phase, into
domains of different orientations.  This phenomenon will be addressed in
Section IV.

\subsection{The TFG and TDS methods}
Free-growth experiments (Fig.  \ref{sketches}a) were performed in an 
Instec hot stage.  The thermal stability of the setup is of a few mK. 
We chose a relatively slow cooling rate (about $0.01~{\rm{Ks^{-1}}}$) 
in order to prevent the system to overshoot the desired value of the 
undercooling $\Delta T = T_{NS} - T$, where $T$ is the targeted 
temperature.  The SmB phase appears by heterogeneous nucleation, at a 
rate which depends on $\Delta T$.  No nucleation events are observed 
within several tenths of minutes for $\Delta T <$ 0.1 K. Nucleation 
rarely occurred for $\Delta T $ between 0.1 and about 0.3 K in 
PI-coated samples.  Within that $\Delta T$ range, the number of 
crystals appearing in the field of view of our optical setup ($625 
\times 480 \mu m^{2}$) does not exceeds two in PTFE-coated samples.  
For the values of $\Delta T$ that we used generally (0.3--0.6 K), two 
to six crystals nucleate within a few seconds in an area of 
$10^{-1}~\rm{mm^2}$.  Unfortunately, nucleation occurs during the 
thermal transient of the hot stage, so that the uncertainty on the 
value of $\Delta T$ is relatively large (0.05 K), which prevented us 
to perform a systematic study of the nucleation rate as a function of 
$\Delta T$ \cite{noteclausius}.

A detailed description of our TDS setup (Fig.  \ref{sketches}b) can be
found in ref.  \cite {akafaihl}.  We used values of $V$ ranging from 1 to
$30~\mu \rm{ms^{-1}}$.  The value of $G$ (from 30 to 80 K{$\rm{cm^{-1}}$})
remained constant within less than $10 \%$ during a given solidification
run.  We used both PI- and PTFE-coated samples in TDS experiments, but most
of the results shown here concern PTFE-coated samples.  A typical TDS
experiment is performed as follows.  A thin sample of CCH4 is introduced in
the solidification setup.  It then melts partly.  The non-melted part of
the sample is a polycrystal, but a large single crystal can be grown from
it using funnel-shaped samples \cite {akafaihl,boak01}.  After a certain
time (about 30 minutes) of maintain at rest ($V = 0$) in order to
homogenize the liquid, the solidification is started at a given velocity.
A stationary regime is generally reached after a transient regime.  Then,
we apply one or several velocity changes and observe the response of the
system to these changes.

Free growth and directional solidification were observed under the
eye-piece of an optical microscope (Leica), either in the bright-field
mode, or using rotating polars.  Images were detected via a CCD camera
coupled to a digital image processing device.  Between crossed polars, a
well-aligned planar nematic appears dark when, and only when, one of the
two polars is parallel to the average direction of the director.  As the
optical axis of the SmB phase is normal to the smectic layers, a
homeotropic SmB crystal appears always dark between crossed polars.  On the
other hand, the contrast between a planarly oriented SmB crystal and the
surrounding planar nematic (or between two SmB grains of different
orientations) depends on the disorientation angle $\phi$.  In principle,
this yields a method for measuring $\phi$ for each grain of a polycrystal.
In fact, because of the recrystallization process which destroys the
initial SmB grain distribution (see below), such measurements had to be
performed during growth.

\section{Growth dynamics of CCH4 smectic B crystals}
\subsection{Free growth}
In Sections IIIA and IIIB, we summarize some TFG and TDS results obtained
in previous studies.  Though, in the present study, we will consider only
planarly oriented crystals, it is useful to recall first that it is
possible, with a suitable surface coating, to obtain thin samples of CCHm
with a homeotropic orientation of the nematic and the SmB crystals.  The
smectic layers are then parallel to the sample plane.  Homeotropic crystals
of CCHm grow non-faceted, thus according to a fully diffusion controlled
dynamics, and 2D dendritic patterns are observed \cite{review}, which
exhibit a six-fold symmetry.  This is consistent with the hexagonal packing
of the molecules within the smectic layers, and typical of a system with a
small value of the interfacial anisotropy.  If the homeotropic SmB crystal
is surrounded by a planar nematic, this introduces an additional (two-fold)
term in the anisotropy (including that of the diffusion coefficient) which
modifies the selection of the dendritic pattern.

Returning to planarly oriented crystals, we note that their orientation is
fully specified by two angles, namely, the disorientation angle $\phi$
defined above (Fig.  \ref{sticks}) and an angle $\alpha$ specifying the
orientation of the hexagonal lattice with respect to the normal {$\bf{y}$}
to the sample plane.  As the dependence of the interfacial properties on
$\alpha$ is very weak --as proven by the observations performed in
homeotropic samples-- the dynamics of the nematic-SmB front for planarly
oriented crystals does not depend sensitively on the angle $\alpha$, and we
will generally ignore it.

\begin{figure}[htbp]
\includegraphics[width=6cm]{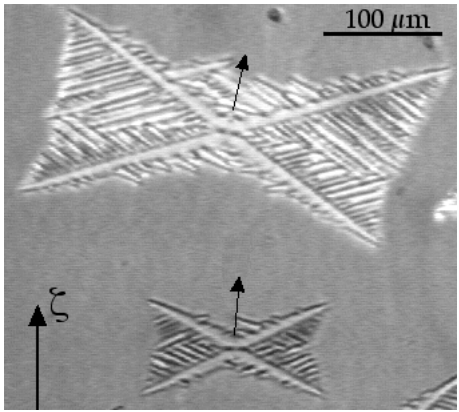}
\caption{Two SmB crystals of CCH4 growing in TFG in a $10$-$\mu$m
thick PI-coated sample ($\Delta T = 0.4$ K).  $\zeta$: rubbing axis of
the PI film.  The snapshot was taken between polars.  Note that no
defects in the nematic alignment are visible.  Small arrows: vector
$\bf{n}_{sm}$ normal to the smectic layers for each crystal (see
text).}
\label{dend}
\end{figure}

A planarly oriented SmB crystal maintained near equilibrium (or 
growing at low undercooling) exhibits an elongated shape, with two 
long facets perpendicular to the sample plane, and rounded ends.  At 
low undercooling (less than about $0.1 $ K), the facets do not grow 
("blocked" facet), within experimental resolution, whereas the rounded 
ends progress with a growth rate less than $10 \mu \rm{ms^{-1}}$ (Fig.  
\ref{sticks}a) \cite {boak01}.  For values of $\Delta T$ slightly 
higher than $0.1 $ K, the facets grow more or less smoothly.  The 
existence of a threshold value of $\Delta T$ below which the facets do 
not grow signals that there is no active lattice defect (dislocations) 
intersecting the interface \cite{bcf}.  Thus the growth of the facets 
for $\Delta T > 0.1 $ K must involve a mechanism of nucleation of 
molecular terraces.  A planarly oriented crystal growing at an 
undercooling larger than $0.1 $ K systematically undergoes shape 
instabilities due to impurity diffusion.  For $\Delta T$ values 
between 0.1 and $0.3 $ K, a mere splitting of the tip (Fig.  
\ref{sticks}b) occurs.  For $\Delta T > 0.3 $ K, dendritic-like 
patterns are observed (Figure \ref{dend}).

In the plane of the sample, the surface tension $\gamma_{NS}$ of the
nematic-SmB interface is highly anisotropic.  The facet orientation
corresponds to a singularity of the Wulff plot, i.e., the angular
dependence of the surface tension $\gamma_{NS}(\bf{n})$ ({$\bf{n}$}: normal
to the nematic-SmB interface) of the nematic-SmB interface, when {$\bf{n}$}
is parallel to the normal ${\bf{n}}_{sm}$ to the smectic layers.  We did
not notice any sign of the existence of forbidden orientations in the
equilibrium shape.  As concerns the interfacial kinetics, it is well
represented by an anisotropic linear kinetic coefficient $\beta (\bf{n})$
defined by $v_{n}=\beta ({\bf{n}}) \Delta T_{k}$, where $v_{n}$ is the
(local) normal velocity of the interface and $\Delta T_{k}$ the kinetic
undercooling, for all orientations except in the close vicinity of
${\bf{n}}_{sm}$.  Realistic $\gamma_{NS}(\bf{n})$ and $\beta (\bf{n})$
functions have been build previously, which account for the main features
of the growth phenomena observed in CCH4 \cite{boto00,boak01}.  However,
the respective contributions of the anisotropies of the surface tension and
of the kinetic coefficient to the selection of the observed growth shapes
are not known with precision.  Either of them, taken separately, could
explain the observations.  We also note that, if $\phi \ne 0$, the nematic
is distorted over a distance comparable to the persistence length in a
region surrounding the crystal, since different anchoring orientations are
imposed along the SmB-nematic interface and along the glass plates.  The
elastic energy associated to that distortion increases obviously with the
disorientation of the crystal.  This may play some role in nucleation and
growth phenomena, as we will see later on.

When the growing crystals exhibit large facets, the disorientation angle
$\phi$ can be measured directly from the micrographs (Fig.  \ref{sticks}).
When the facets are hidden by the development of a dendritic pattern,
$\phi$ can also be estimated (within 1 or 2{$\rm{^o}$}), since
${\bf{n}}_{sm}$ is parallel to the line bisecting the largest angle between
two main dendritic arms (Fig.  \ref{dend}).  This was checked by melting
partly a dendritic crystal and letting it coarsen at a constant temperature
until it reaches a faceted shape.

\subsection{Directional solidification}

The TDS method has been used extensively for the study of nonfaceted
growth \cite{Jackson66,MS,AkaFa98,
LoShiCum98,akafaihl,crho93,karmacta}, but rarely for that of faceted
growth.  In general, faceted crystals exhibit many facets, the growth
of which occurs far from equilibrium and is very sensitive to the
structure of the interface on a molecular scale and to lattice defects
\cite{bcf}.  As a consequence, their (non-local) growth dynamics is
non-stationary on a macroscopic scale \cite{shhu91,fatr97}.  A major
advantage of lamellar phases, e.g., a SmB phase, presenting a single
facet orientation is that stationary or, at least, permanent regimes
can be observed in TDS as a function of the orientation of the
crystals \cite{osme89,meos91,boak01}.

In TDS, the growth dynamics of a planarly oriented single crystal depends
on the orientation of the facet plane with respect to the solidification
axis, i.e., on the angle $\theta$ between the axis {$\bf{z}$} of the
thermal gradient and ${\bf{n}}_{sm}$ (Fig.  \ref{planarfront}) --note that
$\theta = \phi + \phi_{z\zeta}$, where $\phi_{z\zeta}$ is the angle between
{$\bf{z}$} and the direction $\zeta$ of friction \cite{boak01}.  Very
special, non-stationary patterns (not described here) are observed for
$\theta$ within a few degrees of 0{$\rm{^o}$} or 90{$\rm{^o}$}.  For all
the other values of $\theta$, there exist stationary and permanent
patterns, the qualitative features of which are independent of $\theta$.

\begin{figure}[htbp]
\includegraphics[width=7cm]{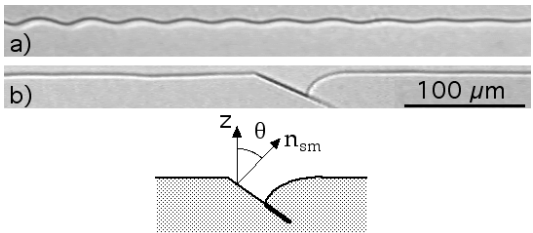}
\caption{Thin-sample directional solidification of CCH4 in a PTFE-coated
sample ($G = 54 ~{ \rm{Kcm^{-1}}}$).  a) Drifting shallow cells ($V= 3.1
~\mu{\rm{ms^{-1}}}$); c) Stationary faceton ($V= 3.1~\mu{\rm{ms^{-1}}}$).
Sketch: definition of the angle $\theta$ (see text).}
\label{cellfaceton}
\end{figure}

\begin{figure}[htbp]
\includegraphics[width=7cm]{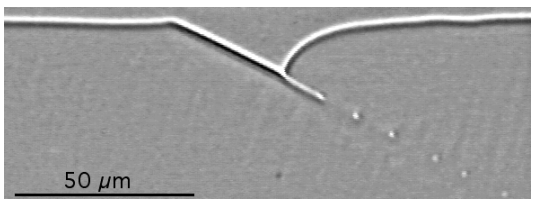}
\caption{The "droplet instability" of the thin groove of a drifting faceton
(TDS; $G = 54 ~{ \rm{Kcm^{-1}}}$; $V= 3.1 ~\mu{\rm{ms^{-1}}}$).}
\label{droplets}
\end{figure}

At rest ($V=0$), the SmB-nematic interface of a single crystal is 
fully nonfaceted, and sits at a $z$ position corresponding to 
$T_{NS}$.  For $V$ below the cellular threshold velocity $V_c$, the 
system reaches the stationary planar-front regime after a certain 
transient time (solute redistribution transient).  The front then 
remains fully nonfaceted (Fig.  \ref{planarfront}).  For $V>V_c$ ($V_c 
\approx 2.5 \mu{ \rm{ms^{-1}}}$ for $G = 54~{\rm{Kcm^{-1}}}$), two 
kinds of patterns are observed, depending on boundary conditions, 
namely, non-faceted patterns made of drifting shallow cells (Fig.  
\ref{cellfaceton}a), and localized, dynamical objects, similar to 
traveling waves, called facetons because their existence is 
intrinsically bound to the presence of a facet which grows at a 
velocity $v_{n}$ generally much smaller than $0.1V$ (Fig.  
\ref{cellfaceton}b).  A faceton appears in most cases from a 
drifting-cell pattern, the amplitude of which is sufficient for a 
small portion of the smectic layer plane to be exposed to the nematic.  
In a faceton, a clearly visible facet extends relatively deeply into 
the crystal.  A thin nematic channel thus forms, which is necessarily 
faceted on both sides.  A faceton either drifts at constant velocity 
along the front ("stationary" faceton), or oscillates while drifting.  
That oscillation corresponds to a relaxation cycle of the facet 
between its blocked state and a state where it is growing.  The 
regularity of the phenomenon inclines us to think that an instability 
of the nematic groove similar to the "droplet instability" observed in 
ordinary deep-cell patterns \cite{kurow}, is at the origin of that 
oscillation (Figure \ref{droplets}).

Since a stringent requirement for observing steady TDS patterns is the
selection of large, planarly oriented single SmB crystals of well controlled
in-plane orientation, we performed most of the TDS experiments in a series
of PTFE-coated samples with different orientations of the friction axis,
i.e., with different values of $\phi_{z\zeta}$.  A counterpart of the use
of PTFE-coated samples is that some defects in the nematic provoke a
permanent perturbation of the SmB-nematic front, even for $V < V_c$
\cite{boak01}.  The major perturbation comes from the nematic-domain
structure (see below).  The fluctuations of the front are clearly slaved to
the domain walls, which remain unperturbed even in the vicinity of the
moving interface, and impose the scale of the perturbation (typically in
the 100-$\mu$m range).  More localized perturbations of the front (on a
scale of the order of 10 $\mu$m) are due to individual defects in the
nematic, most probably disclinations.  These defects are mobile, contrary
to domain walls.  They are generally not destroyed when meeting the front,
but rather migrate along the interface (in the manner of a dust particle)
until they collapse with another defect of opposite sign.

\section {Results}

\subsection {Nucleation, isothermal growth and recrystallization process}

\subsubsection {PI-coated samples}

\begin{figure}[htbp]
\includegraphics[width=8cm]{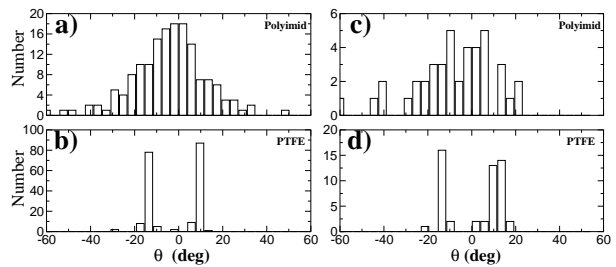}
\caption{Histograms of the final values of the disorientation angle $\phi$
of SmB crystals nucleated in PI-coated samples ({\bf a} and {\bf c}) and in
PTFE-coated samples ({\bf b} and {\bf d}), in TFG ({\bf a} and {\bf b}) and
in TDS ({\bf c} and {\bf d}).}
\label{histograms}
\end{figure}

We measured the disorientation angle $\phi$ for about 200 SmB crystals 
nucleated in two different PI-coated samples.  The $\phi$ 
distribution, shown in the histogram of Figure \ref{histograms}a, is a 
broad symmetric peak centered onto zero.  This is in full agreement 
with previous results \cite {tobo96}.  The same qualitative features 
were also observed for CCH3 and CCH5, but the width of the $\phi$ 
distribution was much narrower (resp., broader) for CCH3 (resp., CCH5) 
than for CCH4.  A similar distribution (Figure \ref{histograms}c) is 
observed in TDS when SmB crystals nucleate ahead of the front (see 
below).

\begin{figure}[htbp]
\includegraphics[width=7cm]{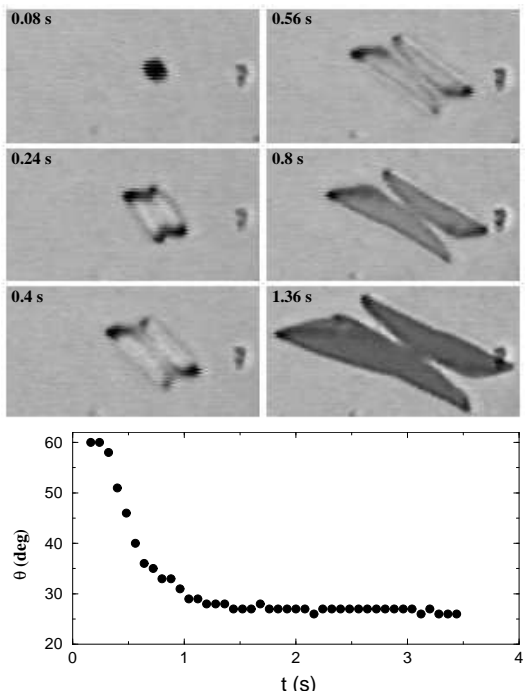}
\caption{Rotation of a highly misoriented SmB crystal of CCH4
nucleating and growing in a PI-coated sample.  Top: Successive stages
of the growth of the crystal.  Horizontal dimension of each snapshot:
$120 \mu m$.  Bottom: Graph of the disorientation angle $\phi$ of the
crystal as a function of time $t$.  The first point was measured when
the size of the crystal was of about $10 \mu m$.}
\label{rotation}
\end{figure}

The values of $\phi$ reported in Fig.  \ref{histograms}a were measured in
well-developed crystals.  In fact, the final orientation of a large crystal
is often different from that of the initial nucleus, i.e., the orientation
of the crystal changes during growth.  Figure \ref{rotation}a shows
successive stages of the growth of a SmB crystal with a large initial
disorientation ($\phi \approx 60 \rm{^o}$).  A plot of $\phi$ as a function
of time $t$ (Fig.  \ref{rotation}b) reveals that the crystal starts
rotating after a delay time of a few 0.1 s, its characteristic dimension
being then of about 10~$\mu \rm{m}$.  It stops rotating (but not growing)
when its (largest) dimension is about 80~$\mu \rm{m}$.  The whole process
occurs within about 1.5 s.  This phenomenon may be explained as follows.
As long as $\phi \ne 0$, an elastic torque is applied to the crystal
because of the distortion of the nematic around it,
whence its rotation motion.  The existence of a delay for the rotation
shows that the nucleus sticks initially to one of the sample walls.  The
final $\phi$ value, which is far from being zero, corresponds to the time
at which the crystal fills the thickness of the sample.  Interestingly
enough, the alignment of such a crystal along $\zeta$ can be completed by
remelting it partly so that it reaches a typical size less than $10~\mu
{\rm m}$.

The rotation of the crystal (analog to that of a damped torsion pendulum)
results from a combination of elastic forces, viscous flow and inertial
forces in a time dependent geometry.  This complex problem is not addressed
here.  We simply note that the thin nematic layers squeezed between the
crystal and the glass plates are extremely distorted, while the distortion
near the other faces of the crystal must be smoother.  Therefore, it may be
conjectured that elastic forces, but also friction forces, strongly depend
on the shape, i.e., on the aspect ratio, of the growing crystal.

The above observations (the broadness of the disorientation distribution,
and the rotation of highly disoriented crystals during growth) show that,
in PI-coated samples, the orientation of SmB crystals is not principally
determined by specific interactions with the polymer film, but by elastic
interactions with the surrounding nematic.  This clearly explains why the
orientation effect is weak.  Probably, the planar orientation is determined
by the same mechanism.  Apparently, the planar-orienting effect is more
efficient than the in-plane one (the reason for this remains unclear), but
SmB crystals with an imperfect planar alignment are also occasionally
observed (${\bf{n}}_{sm}$ is then tilted with respect to the sample plane).

\begin{figure}[htbp]
\includegraphics[width=5cm]{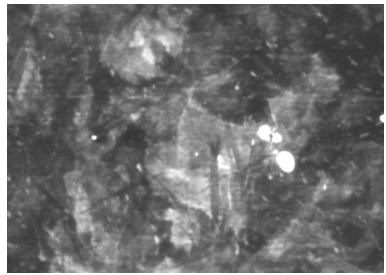}
\caption{Nematic phase in a PI-coated sample initially maintained at room
temperature (it was then a SmB polycrystal) several weeks and heated up
again to $T > T_{NS}$.  The rubbing axis $\zeta$ is horizontal.  Crossed
polars.  The optical contrast due to a memory effect in the nematic was
much enhanced numerically, and corresponds to very slight variations of the
orientation of the nematic director.  Horizontal size: 420~$\mu \rm{m}$.}
\label{memoryPI}
\end{figure}

We gain more information on the interactions between the CCH4 molecules and
the PI film by observing the structure of the nematic in a PI-coated sample
after re-melting the SmB polycrystal.  By melting a fully crystallized
PI-coated sample a short time after a first solidification, the nematic
phase recovers a uniform alignment, even after several crystallization runs
(see the nematic surrounding the SmB Crystal of Fig.~\ref{rotation}).  A
"memory effect" is observed only when the sample is maintained in the SmB
state (at room temperature) several weeks long: the nematic is then
structured into domains of different alignment and containing many defects
(Figure \ref{memoryPI}).  However, the correspondence between the grain
structure of the initial SmB polycrystal and the nematic domain structure
is not clear.  The misalignment angle within each domain is small: the main
alignment effect remains that of the roughness of the PI film.  The
existence of a memory effect evidences the existence of an adsorption layer
of CCH4 onto the PI film.  However, in ordinary experimental conditions,
this adsorption layer, which is probably disordered, has but a weak effect
on the crystallization process.  It will be seen presently that the
situation is completely different in PTFE-coated samples.

\subsubsection{PTFE-coated samples}
\begin{figure}[htbp]
\includegraphics[width=6cm]{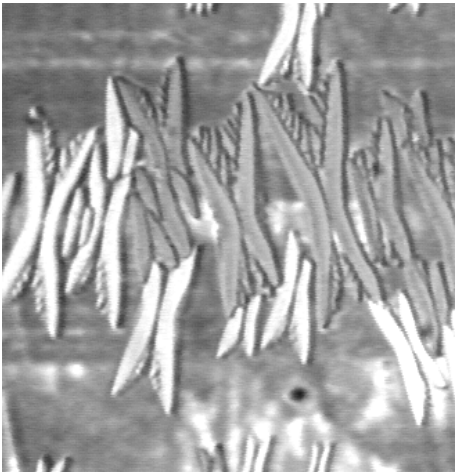}
\caption{Thin-sample free growth ($\Delta T \approx 0.4 $ K).  Smectic B
crystals of CCH4 growing in a $12\rm{-}\mu m$ thick PTFE-coated sample
filled {\it {in situ}} (method 1).  The nematic phase is aligned uniformly
along the horizontal friction axis, except for some defects appearing as
straight thin lines.  Horizontal dimension: 860 $\mu$m.}
\label{nuclPTFE}
\end{figure}

In PTFE-coated samples, nucleation is observed for relatively small 
values of the undercooling ($> 0.1$ K).  Accordingly, the density of 
SmB nucleation sites is larger than in PI-coated ones, for a given 
undercooling.  The SmB crystals are closer to each other, and the 
dendritic patterns are smaller and less branched (Figure 
\ref{nuclPTFE}) than, but similar in shape to crystals observed in 
PI-coated samples.  On the other hand, the crystals are systematically 
tilted with respect to the average nematic orientation $\zeta$ with 
reproducible disorientation angles $\pm \phi_{ptfe}$, where 
$\phi_{ptfe} = 13 \pm 1 \rm{^o}$.  Positive and negative $\phi$ values 
are observed in equal number.  The tails of the distribution (Fig.  
\ref{histograms}b) are essentially due to crystals nucleated onto 
isolated defects of the PTFE films (we did not take the crystals 
nucleated in highly perturbed regions into account).  Thus, just after 
the completion of the solidification, a SmB polycrystal of CCH4 in a 
PTFE-coated sample contains a large number of grains with 
disorientation angles of $\pm \phi_{ptfe}$, and a few grains of 
arbitrary orientations.  Again, a similar distribution is observed in 
TDS (Fig.  \ref{histograms}d).  We return to this phenomenon in Sec.  
IVB.

\begin{figure}[htbp]
\includegraphics[width=7cm]{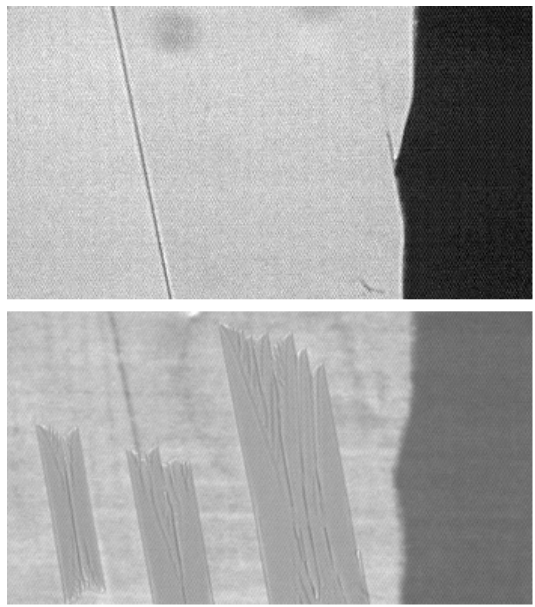}
\caption{PTFE-coated sample of CCH4.  Crossed polars.  The friction axis
$\zeta$ is vertical.  a) Fully crystallized smectic B (TFG; $\Delta T
\approx 0.2 $ K).  The three crystals visible in the image are larger than
the field of view.  The two crystals on the left part of the image have
nearly the same in-plane orientation; b) same sample first heated above
$T_{NS}$, and then cooled down again below $T_{NS}$.  The nematic domains
roughly coincide with the previous SmB grains.  A few SmB crystals which
renucleated during the cooling are also visible.  Horizontal dimension: 570
$\mu$m.}
\label{memory}
\end{figure}

The memory effect is much stronger in re-melted PTFE-coated samples 
than in PI-coated ones.  By re-melting a once solidified PTFE-coated 
sample of CCH4 (Figure \ref{memory}a), one obtains a planar nematic 
phase, which is now structured into domains (Fig.  \ref{memory}b).  
Each nematic domain appears uniformly oriented between crossed polars.  
The nematic orientations differ from one domain to another, and the 
angle $\phi_{nd}$ (where $nd$ stands for nematic domain) between 
$\zeta$ and {$\bf{d}$}, which was initially equal to zero, takes on 
values intermediate between zero and $\pm \phi_{ptfe}$ (we measured 
values between $5\rm{^o}$ and $10\rm{^o}$).  The domains are separated 
from each other by sharp (within thermal fluctuations) boundaries, 
which more or less coincide with the GB's of the polycrystal.  When 
such a "marked" sample is cooled down again to a temperature below 
$T_{NS}$, SmB crystals nucleating within a given nematic domain are 
all of the same orientation (Fig.  \ref{memory}b).  The corresponding 
value of $\phi$ is close (within 1{$\rm{^o}$}) to that of the 
previously grown SmB crystal.  These facts explain the existence of a 
nematic-domain structure in PTFE-coated samples prepared by method 2.

The history-dependent domain structure of the nematic does not disappear
when $T$ is increased to values slightly higher than $T_{NS}$.  This
signals that CCH4 molecules strongly adsorb onto the PTFE film.  The
strength of the adsorption is evidenced by the fact that nematic domains
re-appear after the sample has been maintained overnight at $T \approx 90
\rm{^o}$C, thus in the isotropic state, and cooled down again below
$T_{IN}$ (but slightly above $T_{NS}$).  However, the domain boundaries
appear then much blurred, and the nematic alignment is no more uniform
within a given domain.  Moreover, it seems that the value of $\phi_{nd}$
within each domain is somewhat closer to zero.

The above observations strongly suggest that the epitaxy process at play in
PTFE-coated samples of CCH4 does not occur directly onto the PTFE film
itself, but onto a layer of CCH4 molecules, which has adsorbed when the
nematic first entered into contact with the PTFE film.  This is a case of
homoepitaxy of SmB crystals onto a crystalline layer of CCH4 molecules, the
structure of which is not necessarily that of a stable bulk phase (a
similar phenomenon was observed in thin films of another liquid-crystal
molecule, called 8CB, classically used as a model mesogenic system,
deposited onto the flat surface of a {$\rm{MoS_2}$} single crystal, at a
temperature close to that of the nematic-smectic A transition of the bulk
8CB \cite{lacaze}).  The homoepitaxy phenomenon does definitely not exist
with PI. The two degenerate planar orientations $+\phi_{ptfe}$ and
$-\phi_{ptfe}$ of the macroscopic SmB crystals in PTFE-coated samples must
result from a specific lattice matching between the thin adsorbed layer and
the bulk SmB phase occurring for those values of $\phi$.  That the adsorbed
layer does not determine the alignment of the nematic in freshly filled
samples suggests that the crystalline layer is made of a large number of
very small grains of different orientations, or contains a large number of
defects.  It may also be conjectured that those defects could be induced by
the structure of the PTFE film, that is, either by an irregular topography
on a microscopic scale, or by defects specific of the helix structure of
the PTFE chains.  As long as the adsorbed layer is disordered on a scale
comparable to that of the fluctuations of the nematic order, the bulk
nematic is insensitive to it and is aligned along an average direction
imposed by the microscopic roughness of the film, which is a symmetry axis
of the system.  As the growth of bulk SmB crystals occurs, the adsorbed
layer undergoes a reorganization over long distances, which breaks the
initial axial symmetry about the direction of friction, and modifies the
anchoring of the nematic.  The fact that the nematic appears uniform within
each domain signals that the upper and the lower surface layers (adsorbed
on the upper and the lower glass walls of the sample) are identically
reorganized.  The ordered structure of the surface layers is not much
perturbed after re-melting, as evidenced by the strong memory effect.  The
fact that $\vert \phi_{nd} \vert$ takes intermediate values between
$\phi_{ptfe}$ and zero may be the sign of a competition, in the nematic
alignment effect, between the roughness of the PTFE film and the order of
the adsorbed layer.  When the sample is heated up to the isotropic phase,
either a slow desorption of the molecules occurs, or, more probably, the
adsorbed layer only undergoes a slow disordering.

\subsubsection{Recrystallization process}

\begin{figure}[htbp]
\includegraphics[width=8cm]{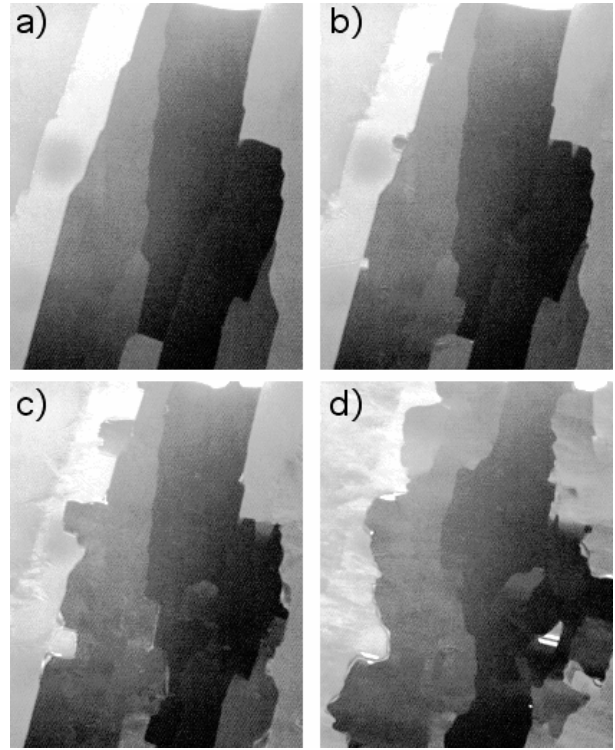}
\caption{Recrystallization phenomenon in a PTFE-coated sample of CCH4 (TFG;
$\Delta T \approx 0.4 $ K).  The friction axis $\zeta$ is vertical.
Crossed polars.  a) $t = 0$ (the crystallization is complete); b) $t = 4
s$; c) $t = 7 s$; d) $t = 48 s$.  Horizontal dimension of each snapshot: 370
$\mu$m.}
\label{recrystal}
\end{figure}

In a PTFE-coated sample, the grain structure of a polycrystal sample
grown in TFG and maintained at a $\Delta T$ value smaller than $0.3 $
K (the SmB grains are then in a small number, thus of large size) does
not evolve in time --such was the case in Fig.  \ref{memory}.  For
$\Delta T > \Delta T_{recryst} \approx 0.3 $ K, the polycrystal
undergoes a recrystallization process, during which some grain
boundaries, or some parts of them, migrate, generally in a stepwise
manner (Figure \ref{recrystal}).  The normal velocity of a grain
boundary can reach a few $10 \mu \rm{ms^{-1}}$.  A transient
three-dimensional (3D) deformation of the GB is sometimes observed,
which evidences a marked sensitivity to the roughness or chemical
heterogeneities of the substrate, like in a wetting process.  The
process is rapid during the first ten seconds, and then slows down.
It is essentially completed within 1 min.

\begin{figure}[htbp]
\includegraphics[width=8cm]{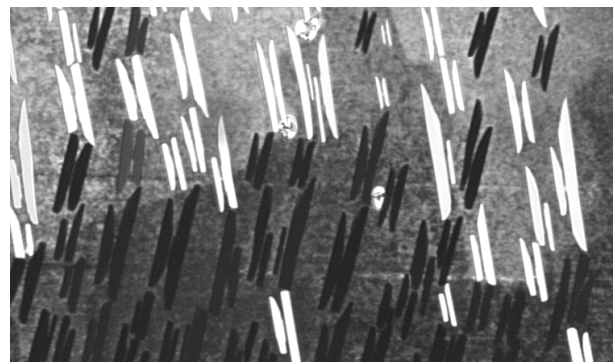}
\caption{Nucleation of SmB crystals in a PTFE-coated sample of CCH4
(TFG; $\Delta T \approx 0.4 $ K).  The nematic-domain structure is
inherited from a first solidification and an isothermal
recrystallization process in the SmB state.  The friction axis $\zeta$
is vertical.  Crossed polars.  Horizontal dimension: 1.2 mm.}
\label{nucl-recryst}
\end{figure}

The domain structure of the nematic phase after melting of a
recrystallized sample keeps memory of both the SmB grain structures
before and after the recrystallization process, even after a long stay
at an undercooling larger than $\Delta T_{recryst}$, and becomes very
complex.  This probably means that the recrystallization process
affects only one of the two (upper and lower) adsorbed layers.
Therefore, after a recrystallization process, the two inner surfaces
of the sample are no longer identical, and the orientation within
nematic domains does not simply reflect that of the adsorbed layers.
This is evidenced by the fact that both $+\phi_{ptfe}$ and
-$\phi_{ptfe}$ disorientation angles of SmB crystals nucleated in a
remelted (once recrystallized) sample are observed within a given
nematic domain (Figure \ref{nucl-recryst}).

In TDS, the recrystallization process takes the form of a "second front"
following the nematic-SmB front at fixed distance corresponding roughly to
$\Delta T_{recryst}$ (Fig.  \ref{planarfront}; also see \ref{obtuse}
below).  A more intriguing configuration, in which the recrystallization
front bends itself to join the nematic-SmB front (see Figure \ref{facetGB}a
below) is also frequently observed.  Such a configuration is not in
equilibrium, but drifts laterally as a whole in a direction corresponding to
the decrease of the size of the high-temperature grain (i.e., leftwards in
the case of Fig.  \ref{facetGB}a).  These observations can be explained
qualitatively in the frame of a first-order transition scheme.  The two
different "phases" (in the definition of which surface effects must be
included) are in equilibrium at a definite temperature $T_{eq}=T_{NS}-
\Delta T_{recryst}$.  The migration observed in Fig.  \ref{facetGB}a means
that the gain in bulk free energy due to the presence of the
high-temperature phase above $T_{eq}$ is less than the loss due to the
presence of the recrystallization front --in other words, the
high-temperature grain has a subcritical size.  However, the quantitative
details, especially the fact that, in Fig.  \ref{facetGB}a, there is no
measurable temperature difference between the G1-liquid and G2-liquid
interfaces, pose problems.  At present, the question of the nature of the
driving force responsible for the recrystallization process remains open.

\subsection{Faceting and nucleation in directional solidification}
\subsubsection{Nonfaceted and faceted GB grooves at rest ($V= 0$)}
\begin{figure}[htbp]
\includegraphics[width=8cm]{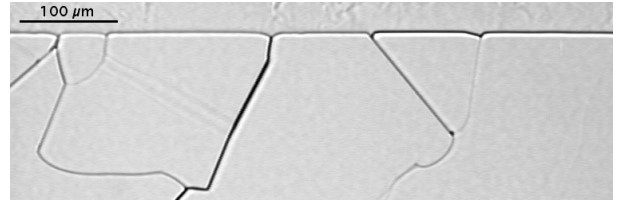}
\caption{The nematic-SmB interface of a CCH4 polycrystal in a 
PTFE-coated sample at rest ($V = 0$) in TDS. End of the coarsening 
process.}
\label{polycrystal}
\end{figure}

In the PTFE-coated samples that we use in directional solidification, most
of the grains exhibit a disorientation angle close to $\pm \phi_{ptfe}$.
Due to nucleation onto defects of the PTFE films, some grains markedly
misoriented with respect to the epitaxy angles are also present.  At rest
in the thermal gradient, the non-melted part of such a polycrystal sample
undergoes a grain coarsening process, which affects the solid far below
$T_{NS}$.  In contrast to the recrystallization process described above,
which occurs below a threshold temperature lower than $T_{NS}$, the
considered coarsening process is particularly active near the solid-liquid
interface.  There, GB's are highly mobile, and rearrange in order to
intersect the solid-liquid interface at right angle.  The motion of the GB's
slows down progressively, and, after several tenths of minutes, GB's have
practically ceased moving (Figure \ref{polycrystal}).  The typical distance
between GB's intersecting the front (grain size) is then of a few $100 \mu
\rm{m}$.  At this stage, the nematic-SmB interface is planar except for a
few shallow grooves (or cusps) attached to GB's.

\begin{figure}[htbp]
\includegraphics[width=5cm]{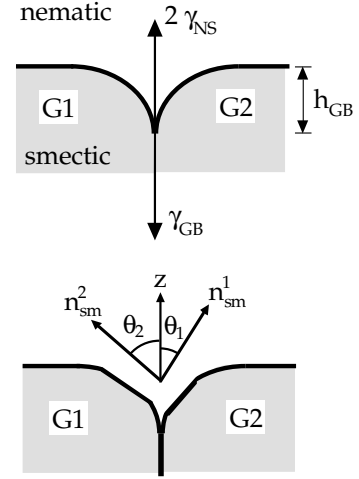}
\caption{Grain boundary grooves (sketches).  Top: nonfaceted groove.
Bottom: faceted groove.}
\label{GBschema}
\end{figure}

In the vicinity of the SmB-nematic interface, GB's run perpendicular 
to the sample plane, and parallel to {$\bf{z}$}.  This (and the high 
mobility of the GB's) shows that the GB's are "wetted" by the nematic 
(an exception to this rule corresponds to GB's running parallel to the 
smectic-layer plane of one of the adjacent grains, indicating a 
singularity of the Wulff-plot of the GB's in that orientation; one of 
such GB's is visible in Fig.  \ref{polycrystal}).  Let $\theta_{1}$ 
and $\theta_{2}$ be the disorientation angles of two adjacent grains 
G1 and G2, respectively (Figure \ref{GBschema}).  The surface tension 
$\gamma_{GB}$ of a GB such that the value of the angle $\theta_{12} = 
\theta_{1} - \theta_{2}$ (which is one of the angular components 
characterizing the misorientation of the GB) is larger than a few 
degrees, is equal to 2$\gamma_{NS}$ \cite{GB}.  The apex angle between 
the two solid-liquid interfaces on the bottom of a GB groove is zero 
(Young's law).  Let us consider the case $\theta_{1} < 0$ and 
$\theta_{2} > 0$.  Then, the smectic-layer plane is not exposed to the 
nematic, and the GB groove is fully nonfaceted.  A rough estimate of 
$\gamma_{NS}$ can be obtained by measuring the depth $h_{GB}$ of the 
groove, and using the fact that, for a wetted GB in an isotropic 
system, $h_{GB}$ is equal to a capillary length $d_c = \sqrt{2a_o/G}$ 
($a_o = \gamma_{NS} T_m/L_v$ is the Gibbs-Thomson coefficient, $T_m$ 
the melting temperature of the pure system, and $L_{v}$ the latent 
heat per unit volume), which is also the length over which the GB 
groove extends along the direction {$\bf{x}$} \cite{SchaGlic75}.  We 
measured directly $h_{GB}\approx 3 \mu \rm{m}$ (within $\pm 1 \mu m$) 
for $G = 54 Kcm^{-1}$, thus $a_{o} \approx 2.5 \times 10^{-8} Km $, 
which gives $\gamma \approx 3 mNm^{-1}$ ($L_v \approx 44 Jcm^{-3}$).  
This is a reasonable value for such a system.

\begin{figure}[htbp]
\includegraphics[width=7cm]{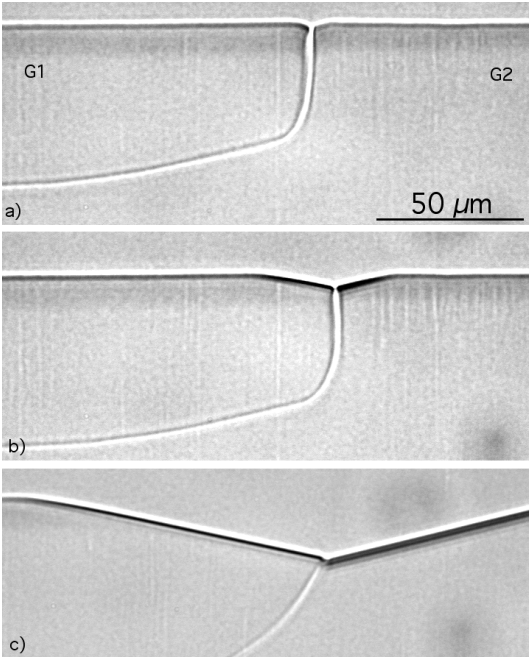}
\caption{A symmetric faceted GB groove in a CCH4 polycrystal in a
PTFE-coated sample in TDS: a) at rest ($V = 0$); b) during the solute
redistribution transient ($V= 2~\mu \rm{ms^{-1}}$); c) V-shaped GB groove
pattern.}
\label{facetGB}
\end{figure}

At rest, a GB groove can be faceted either on both sides if $\theta_1 > 0$
and $\theta_2 < 0$, or on one side only if $\theta_1$ and $\theta_2$ are of
the same sign.  The facets are generally hardly visible when the front is
strictly at rest (Figure \ref{facetGB}a).  On the other hand, if the front
slightly advances, either because of an accidental perturbation of the
thermal field, or of a slow drift of the grain boundary along the front,
the facets appear clearly (Fig.  \ref{polycrystal}).  This is a further
evidence of the fact that the facets remain blocked at small undercoolings.

\subsubsection{Grain selection mechanism}

The drifting motion of asymmetric GB groove patterns is the main
ingredient of a grain selection mechanism at play in PTFE-coated
samples.  It is therefore worth studying the mechanisms of drift of
the GB grooves in some detail.  Several cases corresponding to
different signs of $\theta_{1}$ and $\theta_{2}$ must be considered in
turn.  When a GB groove is nonfaceted ($\theta_{1} < 0$ and
$\theta_{2} > 0$), its shape does not change significatively, and the
large-scale dynamics of the front is not disturbed at low velocity ($V
< V_{c}$).  On the other hand, for $V > V_{c}$, a precursory
deformation of the front in the vicinity of the GB groove serves as an
initiator for the cellular instability, as it is generally observed in
TDS experiments \cite{CoriellSekerka}.

When a GB groove is faceted, it starts to deepen from the onset of the
pulling.  The pre-existing facets extend continually during the solute
redistribution transient (Figs.  \ref{facetGB}b and \ref{facetGB}c).  They
recoil first at a velocity nearly equal to -$V$, i.e., they do not, or
almost not grow.  When the undercooling of the coldest end of the facets
reaches a value of about $0.1 $ K, they start growing, and the deepening of
the GB groove slows down.

\begin{figure}[htbp]
\includegraphics[width=8cm]{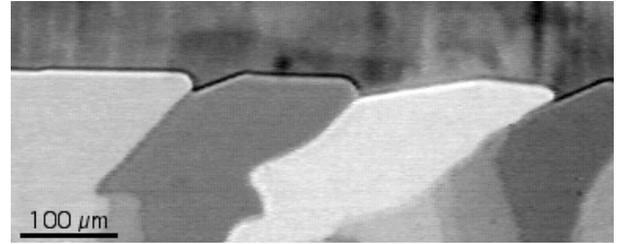}
\caption{Faceted patterns ("GB locked facetons") attached to GB's in TDS
($V=3~\mu \rm{ms^{-1}}$) of a CCH4 polycrystal (PTFE-coated sample).  These
patterns drift along the front, as evidenced by the tilt of the GB's in the
solid.}
\label{lockfaceton}
\end{figure}

\begin{figure}[htbp]
\includegraphics[width=6cm]{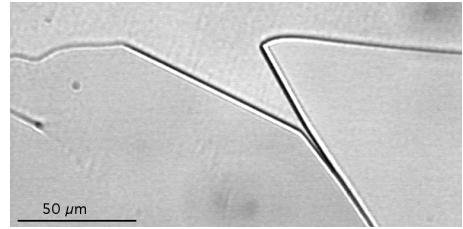}
\caption{A fully faceted (drifting) pattern attached to a GB in TDS of
a CCH4 polycrystal (PTFE-coated sample).}
\label{lockfacetonbig}
\end{figure}

When the GB groove is faceted on one side only ($\theta_{1}$ and
$\theta_{2}$ of the same sign), a localized, permanent pattern forms, which
drifts along the front.  If $V < V_{c}$, the groove remains faceted on one
side only, the drifting motion is governed by that of the facet, and the
non-faceted side of the pattern slightly bulges in the nematic towards the
drifting direction (Figure \ref{lockfaceton}).  The pattern is then very
similar to a faceton locked onto a GB ("GB-locked faceton"), and drifts
laterally at a constant velocity.  A GB formed by this mechanism is tilted
in the solid with an angle which is determined by the drift of the GB
groove pattern.  In some cases, the other facet, that did not form at rest
for geometrical reasons, appears, which results in a pattern such as that
shown in Figure \ref{lockfacetonbig}.

\begin{figure}[htbp]
\includegraphics[width=7cm]{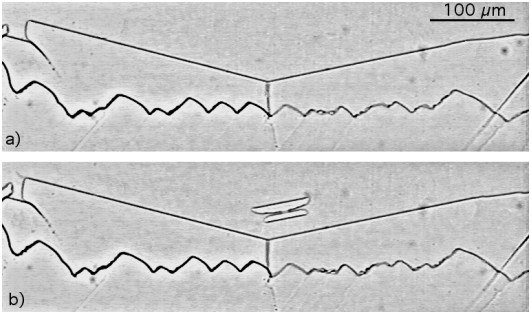}
\caption{Obtuse V-shaped faceted pattern attached to a GB in TDS of a CCH4
polycrystal in a PTFE-coated sample with friction axis parallel to the
solidification one ($V=14 \mu \rm{ms^{-1}}$).  a) Stationary pattern; b)
Nucleation of a new crystal.}
\label{obtuse}
\end{figure}

\begin{figure}[htbp]
\includegraphics[width=8cm]{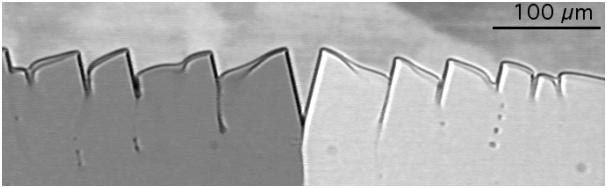}
\caption{Stationary acute V-shaped faceted pattern attached to a GB in TDS
of a CCH4 polycrystal in a PTFE-coated sample with friction axis
perpendicular to the solidification one ($V=14 \mu \rm{ms^{-1}}$).  }
\label{acute}
\end{figure}

The grooves that are faceted on both sides ($\theta_{1} > 0$ and
$\theta_{2} < 0$) can be either symmetrical ($ \theta_{1} =
-\theta_{2}$) or asymmetrical ($\theta_{1} \ne -\theta_{2}$).  The
latter drift laterally, whereas the former do not drift.  In
PTFE-coated samples with the PTFE friction axis $\zeta$ perpendicular
("$\bot$ samples") or parallel ("$\parallel$ samples") to {$\bf{z}$},
most, but not all, crystals are "well oriented", i.e., their in-plane
orientation is such that $\vert \theta \vert$ is close to
$\phi_{ptfe}$ and to $\pi/2 - \phi_{ptfe}$ in $\parallel$ and $\bot$
samples, respectively.  In the first stages of the solidification run,
the drift of asymmetric GB grooves and of GB-locked facetons leads to
the elimination of most of the misoriented grains, while well oriented
grains extend laterally.  This leads to the elimination of the few
grains that have a disorientation angle different from $\pm
\phi_{ptfe}$.  Grains of positive and negative $\theta$ values then
alternate ($\theta_1 \approx -\theta_2$), and are separated either by
non-faceted grooves or by faceted GB grooves with a symmetric shape,
called "V-shaped" patterns (Figures \ref{obtuse}a and \ref{acute}).
The angle between two adjacent facets is equal to about $2\phi_{ptfe}$
($\pi - 2\phi_{ptfe}$) in $\bot$ ($\parallel$) samples.  For $V <
V_{c}$, V-shaped patterns are essentially stationary.  The average
normal velocity of the facets is then $V{\rm{cos}}\theta$.  We will
see later on that the growth of the facets is in fact irregular on a
short time scale.

The depth $h$ of V-shaped patterns does not depend much on $V$ when $V
< V_{c}$.  It is comparable to that of a faceton when $V$ is close to
$V_{c}$, and increases with $V > V_{c}$.  It is typically $60~\mu m$
for $G = 54~Kcm^{-1}$ ($V_{c} \approx 8~\mu{ \rm{ms^{-1}}}$) and $V =
14~\mu{ \rm{ms^{-1}}}$, which corresponds to an undercooling in the
bottom of the pattern of about 0.5 K (for such $V$ values, the whole
front is composed of faceted fingers, typical of the high-velocity
regime \cite{boak01}).  This is potentially sufficient for nucleation
to occur within the nematic trough bordered by the facets.  Nucleation
events are indeed observed in "obtuse" ($ \theta_{1} = -\theta_{2} =
\phi_{ptfe}$) V-shaped grooves (Figure \ref{obtuse}b) in $\parallel$
samples.

The phenomenon of nucleation of crystals ahead of the solidification
front is a common one in directional solidification above the cellular
threshold \cite{meos91}.  The frequency of the nucleation events
depends on the density of nucleation sites $n_{s}$, which is small in
the present system (the time lapse between two successive nucleation
events is of several seconds for $V = 14 \mu ms^{-1}$).  Remarkably
enough, no nucleation events are observed in "acute" ($ \theta_{1} =
-\theta_{2} = \pi /2 - \phi_{ptfe}$) V-shaped grooves ($\bot$
samples).  This is due to the fact that, as nucleation sites active
for $\Delta T$ values smaller than 0.3 K are rare, as shown by TFG
experiments, the extension of the nematic region bordered by the
facets in acute V-shaped troughs is too small (much smaller than in
the obtuse ones) for nucleation to occur.  By estimating the flux $f$
of nucleation sites through a V-shaped pattern as being equal to
$2Vn_{s}h/\rm{tan}(\phi _{ptfe})$ for an obtuse pattern and
$2Vn_{s}h/\rm{tan}(\pi/2 - \phi _{ptfe})$ for an acute one, one finds
that the ratio between the two values of $f$ is equal to
$\rm{tan}(\pi/2 - \phi _{ptfe})/\rm{tan}(\phi _{ptfe}) \approx 20$,
which agrees well with the proposed explanation.

\begin{figure}[htbp]
\includegraphics[width=8cm]{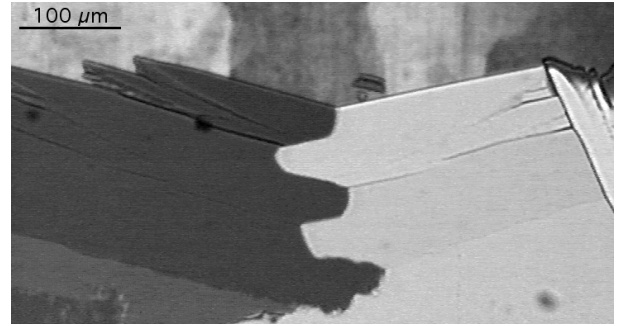}
\caption{Reiterated nucleation events in an obtuse V-shaped faceted
pattern attached to a GB in TDS of a SmB polycrystal in a PTFE-coated
sample of CCH4 with friction axis parallel to the solidification one
($V=14 \mu \rm{ms^{-1}}$).}
\label{cyclenucl}
\end{figure}

In an obtuse V-shaped pattern, each new crystal grows rapidly (within
several $0.1 s$) in conditions approximately similar to a free-growth
configuration with a regularly increasing undercooling.  It thus fills
rapidly the lowest part of the groove.  As expected from observations
in TFG, the disorientation angle of crystals nucleating in V-shaped
troughs is generally $+\phi_{ptfe}$ or $-\phi_{ptfe}$ (Fig.
\ref{histograms}d), according to the nematic domain within which they
appear.  At the end of the nucleation and growth process, the new
crystal is practically undistinguishable from one of the two
pre-existing neighboring grains, and no GB is formed, if $\phi$ is
strictly equal to $\pm \phi_{ptfe}$.  When a GB (or a subboundary)
forms, it is highly asymmetric, and thus drifts rapidly along the
front.  When the new grain meets the previous grain of opposite
orientation, an obtuse V-shaped groove is restored.  This groove
starts then to deepen again, and the whole process can reiterate
cyclically (Figure \ref{cyclenucl}).

Finally, several phenomena --the epitaxy of SmB crystals nucleating during
solidification, the lateral drift of GB's between misoriented grains and the
fact that symmetric patterns associated to well oriented grains are
stationary or cyclically restored-- work towards a grain selection
mechanism in $\parallel$ and $\bot$ PTFE-coated samples.  Only a major
disturbance (e.g., the nucleation of markedly misoriented crystals onto
defects or the meeting of the recrystallization front with the bottom of a
GB groove) may lead to the destruction of the thus selected polycrystal.

\subsubsection{Facet growth and formation of grain subboundaries}

In a polycrystal sample, there are not only GB's of large
misorientation, but also grain subboundaries (SBs).  A SB is made of a
regular arrangement of dislocations.  Its surface tension
$\gamma_{SB}$ is less than 2$\gamma_{NS}$, and decreases when the
misorientation decreases --it is more or less proportional to the
misorientation angle $\theta_{12}$.  Accordingly, the depth $h_{SB}$
of the groove created by a SB emerging at the solid-liquid interface
is less than $d_c$.

The presence of SBs is clearly revealed during a solidification run,
because, as a SB groove slightly deepens, a small facet appears
systematically on one side of it.  The depth of such a SB groove
--thus the size of the facet-- is small (it does not exceed a few $\mu
m$), and the facet remains in a blocked state.  Consequently, SB
grooves drift laterally along the front at a constant speed equal to
$V/\rm{tan}\theta$.  The SB left in the solid is tilted with an angle
equal to $\theta$ and is parallel to the smectic plane of one of the
grains.

We observed that a relatively large number of SB grooves permanently sweep
the front during a long-time solidification run.  As they drift, they are
necessarily eliminated when they meet one edge of the sample (or a GB).
Therefore, there must exist a mechanism of creation of SBs during growth.
We did not observe the polygonization during growth described recently by
Bottin-Rousseau et al \cite{bottin} in TDS of nonfaceted organic crystals.
In the CCH4 system, SBs are emitted from the large facets attached to the
GB's, as it will be explained presently.

\begin{figure}[htbp]
\includegraphics[width=7cm]{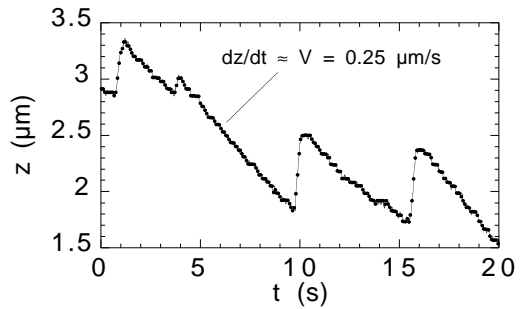}
\caption{The $z$ position of the front at a fixed coordinate $x =
50~\mu{\rm{m}}$ within the facet of grain G1 of Fig.  \ref{facetGB} as
a function of time $t$ at the end of the transient recoil of the
nonfaceted part of the front.}
\label{recul}
\end{figure}

The growth of facets bordering a V-shaped pattern attached to a
symmetric GB occurs in a stepwise manner.  This can be seen by
recording the $z$ position of a point of the facet (at fixed $x$) as a
function of time $t$ (Figure \ref{recul}).  Most of the time, the
facet recoils towards the cold part of the setup at a velocity close
to $V$.  It is thus in a (nearly) blocked state.  At time intervals of
a few seconds (for $V$ in the 1-$ \mu{ \rm{ms^{-1}}}$ range), the
facet seems to progress very rapidly (within much less than $1 s$)
towards the liquid.  In fact, this corresponds to the motion of a
macrostep along the facet.

\begin{figure}[htbp]
\includegraphics[width=8cm]{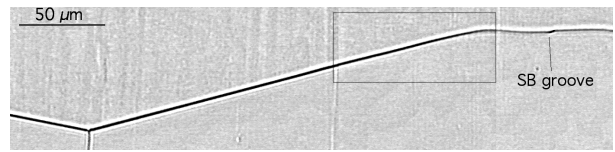}
\caption{Large V-shaped faceted GB groove in a PTFE-coated sample (TDS; $G
= 54~{\rm{Kcm^{-1}}}$; $V = 3~\mu \rm{ms^{-1}}$).  The friction axis
$\zeta$ is parallel to the solidification axis $z$.  A small facet drifting
laterally towards the right side of the sample signals the presence of a
SB. Frame: region analyzed in Fig.  \ref{macrostep}.}
\label{largefacet}
\end{figure}

\begin{figure}[htbp]
\includegraphics[width=8cm]{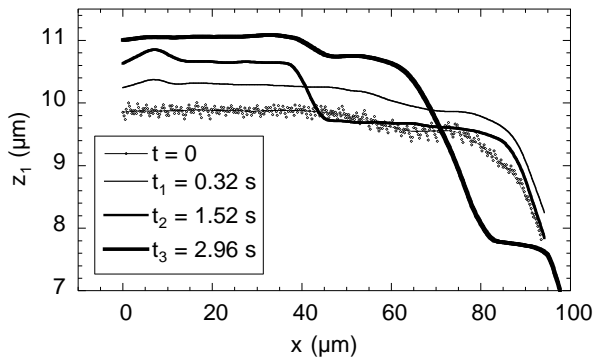}
\caption{Shape $z_{1}(x)$ of the moving nematic-SmB interface in the
vicinity of the edge of the V-shaped pattern of Fig.  \ref{largefacet} at
successive times.  The average slope of the facet has been subtracted to
the $z(x)$ curves.  The curves have been shifted apart from each other by
an arbitrary value for the sake of clarity.  Data points are shown for the
curve at $t = 0$.}
\label{macrostep}
\end{figure}

The process of creation and propagation of macrosteps is illustrated in
Figures \ref{largefacet} and \ref{macrostep}.  In Fig.  \ref{largefacet},
one can see the right part of a large, stationary V-shaped GB groove
pattern.  We have recorded the shape of the SmB-nematic interface in the
region delimited by the frame in that figure as a function of time.  Three
profiles corresponding to successive times are shown in Fig.
\ref{macrostep}.  In that figure, the average slope of the facet is
subtracted from the interface shape, so that local departures from a flat
facet are emphasized.  Time $t = 0$ was chosen at a moment when the facet
was nearly blocked and is approximately flat, except in the region where it
joins the rough part of the SmB-nematic interface.  At time $t_1$, a small
bump appears on the left part of the figure, at some position $x$.  About
one second later (time $t_{2}$), that bump has transformed into a macrostep
of an amplitude of about $1~\mu m$ which propagates along the facet.  The
macrostep changes its shape and increases in amplitude as it progresses,
but the advancing speed of its foremost point is approximately constant.
At time $t_{2}$, a new bump has appeared at the rear of the macrostep, at
the same place as the former one.  When a macrostep reaches the external
edge of the V-shape pattern, thus the planar, rough part of the growth
front, it quite systematically emits a very small drifting facet, such as
that shown in Fig.  \ref{largefacet}.

Our observations give a clear evidence that the main mechanism of
faceted growth in CCH4 is the propagation of steps from terraces
nucleating onto preferential sites, and undergoing a bunching
instability leading to the formation of macrosteps.  Small bumps
appear repetitively as precursors of the macrosteps at one and the
same $x$ position.  We have observed that phenomenon many times.  This
means that preferential sites of terrace nucleation are situated onto
the PTFE film, and are aligned along $\zeta$.  Those sites are
probably of the same nature as the crystal nucleation sites.  The
point that we want to emphasize here is that terrace nucleation
events, which are followed by the appearance of a macrostep, most
probably correspond to cases where nucleation occurs with a slight
misorientation.  In other words, the emission of small facets drifting
along the rough part of the front is the signature of a planar lattice
defect associated to the formation of macrosteps.  Those defects do
not produce any detectable optical contrast when observed between
crossed polars.  They thus may be stacking faults, which are known to
be easily created in a SmB phase, or SBs associated to an out-of-plane
misorientation, i.e., a slight rotation about the normal to the
smectic layers (variation of the angle $\alpha$ defined above), which
is the optical axis of the SmB crystal.  The latter one is the most
plausible one, since the size of the small drifting facets is not a
constant (it depends on the misorientation of the SB).

\section{Conclusion}

We have studied the mechanisms of crystal orientation in solidification
experiments in thin samples of a mesogenic substance, CCH4, which undergoes
a phase transition between a nematic and a smectic B. We have shown that
the nature and the efficiency of those mechanisms depend much on the nature
of the nucleation substrate, namely, a polymer film coating the inner
surface of the glass-wall container.  The use of samples coated with
monooriented PTFE films leads to unexpected phenomena of grain selection
and of generation of lattice defects in thin-sample directional
solidification.  A better understanding of those mechanisms would require
the use of techniques of investigations on a microscopic scale.

\begin{acknowledgments}
We would like to thank \'A. Buka and T. T\'oth-Katona for providing 
the CCH4.  We benefited from fruitful discussions with M. Brunet.  We 
thank G. Faivre for his critical reading of our manuscript.  One of us 
(TB) benefited from a Marie Curie Fellowship of the European Community 
program IMPROVING HUMAN POTENTIAL under contract number 
HPMF-CT-1999-00132.

\end{acknowledgments}


\begin{thebibliography}{99}
\bibitem{pretilt} In fact, it is probable that PI films do not impose a
strictly planar orientation of the nematic CCH4, but that the molecules are
slightly tilted with respect to the wall plane (pretilt angle), as it has
been shown to be the case for mesogenic molecules with cyanobiphenyl head
groups.
\bibitem{gold} We checked that the nematic alignment is preserved if the
PTFE film is covered with an evaporated gold layer of about 500 \AA, i.e.,
when the PTFE chains are isolated from the CCH4 molecules, but the 1D
character of the roughness is almost preserved.
\end{thebibliography}
\end{document}